\documentstyle[epsf]{l-aa}
\begin{document}
\title{ Helioseismology and the solar age}
\author{ W.A.~Dziembowski\inst{1,2},
G.~Fiorentini\inst{3,4}, B.~Ricci\inst{3,4},
\and R. Sienkiewicz\inst{2}}
\institute{Warsaw University Observatory, Al. Ujazdowskie 4, 00-478 Warszawa, 
Poland
\and N. Copernicus Astronomical Center, Polish
           Academy of Sciences, Bartycka 18, 00-716 Warszawa, Poland
\and Dipartimento di Fisica dell'Universit\`a di Ferrara, 
       via Paradiso 12, I-44100 Ferrara, Italy
\and Istituto Nazionale di Fisica Nucleare, Sezione di Ferrara, 
      via Paradiso 12, I-44100 Ferrara, Italy}
\offprints{W.~A.~Dziembowski}
\thesaurus{06.01.1, 06.05.01, 06.09.1, 06.15.1}
\date{received October 1998 ; accepted .........}
\maketitle
\markboth{W.A. Dziembowski et al.: 
          Helioseismology and the solar age}{}
\begin{abstract}

The problem of measuring the solar age by means of helioseismology has been 
recently revisited by Guenther \& Demarque (1997) and by  
Weiss \& Schlattl (1998). Different best values for $t_{\rm seis}$ and 
different assessment of the uncertainty resulted from these two works. 
We show that depending on the way seismic data are 
used, one may obtain  $t_{\rm seis}\approx 4.6$ Gy, close to the 
age of the oldest meteorites, $t_{\rm met}=4.57$ Gy,
like in the first paper, or above 5 Gy like in the second paper.
The discrepancy in the seismic estimates of the solar age may be eliminated by 
assuming higher than the standard metal abundance and/or an upward 
revision of the opacities in the solar radiative 
interior.  

We argue that  the most accurate and robust seismic measure of the solar 
age are the 
small frequency  separations,  $D_{\ell,n}=\nu_{l,n}-\nu_{\ell+1,n-1}$, 
for spherical harmonic degrees $\ell=0,2$ and radial orders $n\gg\ell$.
The seismic age inferred by  minimization
of the sum of squared differences between the model and the solar 
small separations is
$t_{\rm seis}=4.66\pm0.11$, a number consistent with meteoritic data.
Our analysis supports earlier suggestions of using small frequency separations
as stellar age indicators.

\end{abstract} 
\section{Introduction}
The idea that 
helioseismology may be used to test the assumption 
that the solar age is equal to the age the oldest
meteorite is not new. Gough \& Novotny (1990,  who considered the problem
in great detail,  concluded that the accuracy of 0.3 Gy may 
be achieved once the seismic age indicators are measured to 
a precision of 0.1 $\mu$Hz. 
The precision of current seismic data is now significantly better.
However, results of recent studies of the problem yields conflicting 
conclusions.

Before we go to the results of these studies, let us first point out
that we cannot expect a unique determination of the solar age from 
seismic data.
Calculated p-mode frequencies depend on the assumed solar
age but they also depend on other input solar parameters and physical 
quantities. All these data are subject to uncertainties. 
We now have at our disposal nearly 2000 accurate frequency data for 
solar p-modes to determine solar age -- the only observable 
in the standard solar model (SSM) construction  which we surrender. 
It would be indeed surprising 
if the answer would not depend on the way we make use of seismic data. 
An assessment
of the uncertainty of $t_{\rm seis}$ is even more problematic. 

Guenther \& Demarque (1997) concluded their comparison of the solar 
frequencies with those for models calculated upon assuming different age
with the following statement: ``The best agreement with the calculated 
oscillation spectra is achieved for $4.5\pm 0.1$ Gy". 
Unfortunately, they did not explain how these numbers were 
obtained. 

Weiss \& Schlattl (1998), proceeding in a more formal way, 
used $\chi^2$ minimization to determine $t_{\rm seis}$. They considered 
various seismic observables and corresponding parameters in the model 
calculated for various assumed solar ages.
The observables include surface Helium abundance, $Y_{\rm seis}$,
depth of the convective zone, $r_{\rm cz}$, sound speed in the 
the radiative interior, and the radial mode 
frequencies. In nearly all the cases they considered, the minimum was reached 
for age well above 5 Gy. Typical values of $t_{\rm seis}$ they 
derive  are in the range 5.1 -- 5.2 Gy. 
Taken for granted, the high values of the solar age would mean an 
essential revision of our views on the evolution of the solar system. This is 
not what Weiss and Schlattl (1998) propose. Rather, they regard 
the difference between $t_{\rm seis}$ and $t_{\rm met}$ as a measure 
of the uncertainties in the age determination based on 
the state-of-art stellar evolution theory. 

The main motivation for our work was to explain  
the large difference in the conclusions of the two papers regarding 
the value of $t_{\rm seis}$ and its uncertainty.  Weiss and Schlattl (1998)
themselves have addressed this problem but we did not find their 
explanation sufficient. 
We will begin with providing some information about new solar models 
calculated for the purpose of this investigation. In the main part 
of the paper, we review 
the inference about the solar age based on various seismic observables 
and we identify those which we believe are good age indicators.

\section{New solar models}

We constructed a large number of solar models taking into account diffusion 
of Helium and heavy elements following Thoul et al.(1994). 
In one model (Model 5), which we refer to in Section 5 diffusion
was ignored.
In all the models, we use OPAL
equation-of-sate (Rogers et al., 1996). 
For opacity we use the newest Livermore opacity table (OPAL96, Iglesias
\& Rogers, 1996) for Grevesse \& Noels (1993) heavy element mixture. 
For comparison, we calculated one model using an earlier version of 
the Livermore opacities (OPAL92, Iglesias et al., 1992). At low temperatures 
we used Alexander \& Ferguson (1994) data on molecular and grain opacities.
Nuclear reaction rates are calculated according to
Bahcall \& Pinsonneault (1995). We calculated one model 
(Model 4, see Section 5) with modified reaction rates, still within the
range of uncertainties quoted by Bahcall \& Pinsonneault.  

We assumed the value of photospheric radius $R_{\rm ph}$=696.3 Mm. 
This value is
by $0.8$ Mm higher than the most recent determination of Brown \& Christensen
-Dalsgaard (1998). The reason for our choice is a better agreement
with the seismically inferred sound-speed in the lower convective zone.
The small difference is inconsequential for the conclusions of this work. 
The model radii were fitted to the adopted value with the precision 
better than $5\times10^{-5}$. The luminosity was assumed $3.86\times10^{33}$
erg/s and models were fitted  to precision 
better than $2\times10^{-4}$.

We calculated a number of models for various values of the age, $t$, at the 
standard value of the metal-to-hydrogen ratio, $Z/X=0.0245$,
and at an enhanced 
value of 0.027. The parameters for selected models are listed in Table 1.

\begin{table*}
\caption{Parameters of selected solar models}
\begin{flushleft}
\begin{tabular}{cccccccccccc}
\hline
 Model&  $t$ [Gy]&  $Z/X$& OPAL& $Y_0$& $Z_0$& $Y _{\rm ph}$& $Z _{\rm ph}$& 
 $r_{\rm cz}/R_{\rm ph}$ & $X_c$& $\rho_c$[g/cm$^3$] & $T_c \, [10^6\, {\rm K}]$\\
 \hline  
 0& 4.57& 0.0245& 1996& 0.2739& 0.02024& 0.2429& 0.01811&  0.7163& 0.3331& 157.1&   15.803\\
 1& 5.00& 0.0245& 1996& 0.2705& 0.02045& 0.2386& 0.01821&  0.7109& 0.3090& 164.5&   15.927\\
 2& 4.57& 0.0270& 1996& 0.2814& 0.02199& 0.2502& 0.01971&  0.7126& 0.3212& 157.9&   15.934\\
 3& 4.57& 0.0245& 1992& 0.2777& 0.02010& 0.2467& 0.01801&  0.7141& 0.3297& 157.6&   15.841\\
\hline
\end{tabular}
\end{flushleft}

\end{table*}

A comparison between Model 0 and Model 1 shows the effect of the age on 
main parameters of the solar models. The older sun  ($t> t_{met}$) 
has produced a larger amount of
Helium in the  core. Longer evolution implies also a larger effect 
the gravitational settling i.e. a larger
difference between the initial Helium abundance,
$Y_0$, and the present abundance in the outer layers, $Y _{\rm ph}$. 
In order that the
solar model accounts for the same luminosity, one has
to reduce the initial Helium abundance, with respect to that of the SSM.
With the exception of the energy production region, 
the Helium abundance is reduced everywhere in the solar model
and one can thus explain the following features:

\noindent
i)The present photospheric Helium abundance is lower.

\noindent
ii)Matter is more opaque to radiation,
so that convection starts deeper in the sun.

\noindent
iii)Below the convective zone and above
the energy production core, due to the reduced
``mean molecular weight'' $\mu$ the sound speed is higher.

\noindent
iv)In the energy production core, the effect of
Helium accumulation should dominate, resulting
in a larger $\mu$ and consequently a smaller sound
speed.

Of course, the opposite occurs for a younger
sun. In the next section we will discuss in more detail the differences 
in the sound speed between various models.

\section{Inference from seismically determined solar parameters}

Solar age cannot  be directly determined by means of helioseismology.
In all the approaches, including this one, families of solar models with
various assumed ages are calculated and $t_{\rm seis}$ is 
determined by means of a comparison of more direct seismic observables. 
The most direct are the frequencies, but with no additional assumptions
one may use the density, $\rho(r)$, or the squared isothermal sound speed, 
$u(r)$, determined by means of the frequency inversion. These two functions  
are linked by the hydrostatic equilibrium condition. From $u$, 
$\rho$ and their derivatives one may determine a number of other useful 
structural functions. If, in addition, we assume equation of state (EOS) data,
we may infer the values $Y _{\rm ph}$ and $r_{\rm cz}$.
The last two seismic observables were used by  Weiss \& Schlattl (1998) 
in their first
attempt of the solar age determination. They subsequently, considered also
other quantities. There are various possibilities. We regard a comparison of 
the sound speed as most revealing. The value of $r_{\rm cz}$ does not contain 
independent information and, since it is determined from the derivative of $u$,
it is less accurate.

\subsection{The sound speed}

The result of the inversion for $\delta u/u$ -  the relative 
difference in $u(r)$
between the sun and model 0 is shown in Fig.1, where $r=R$ corresponds to
the temperature minimum. 
In the same plot we show the difference 
in $u$ between some other models (see Table 1) and model 0.

\begin{figure}
\epsfxsize=0.99\hsize
\epsffile{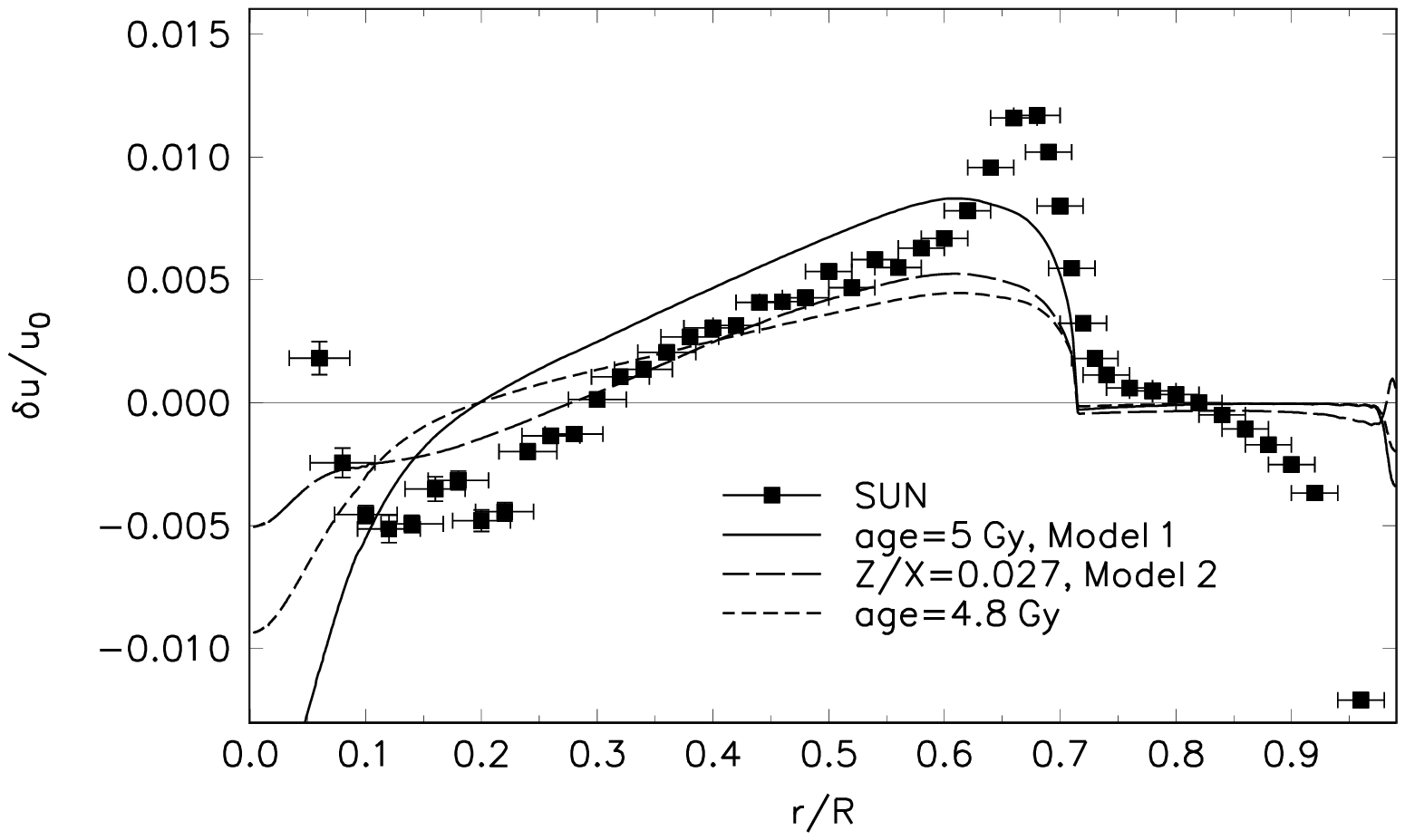}

\noindent Fig.1. Relative differences in $u$ between the sun 
and Model 0 determined by means of helioseismic inversion. 
Also shown are the difference between different models and Model 0.
 The vertical error 
bars (visible only for the inner most points) reflect only measurements
errors. True uncertainty of the inversion is much greater 
(Degl'Innocenti et al., 1997).  
 
\end{figure}

The solar data were obtained from 
the inversion of the frequency data obtained wit the MDI instrument 
(Rhodes et al. 1997) and the GOLF instrument (Gabriel et al. 1997) on board 
of the SOHO spacecraft. 
The first data set contains modes
with the $\ell$ values from 0 to 250. We ignored the f-modes, and we were 
left with the frequencies of 1890 p-modes with $\ell$ up to 184. 
The second set contains
153 frequency data for modes with $\ell$ degrees up to 5. The data were 
combined into a set of 1945 p-mode frequencies. The inversion was done 
by means of the SOLA method (Pijpers \& Thompson, 1992; Dziembowski et al.,
1994). 
  
One sees in Figure 1 that the difference in $u$ through most of the 
sun interior seems to favor higher age. However, the quantitative answer 
depends on the choice of the location in the sun's interior. 
In the region $0.1R < r < 0.35R$,
 $u$ is almost independent of the age. 
In the inner core the dependence on age is the strongest. Older 
models have higher Helium abundance, hence higher mean molecular weight. This
effect dominates the sound speed behavior. 
Unfortunately, results of seismic sounding of the inner core are unreliable.

An assessment of the solar age based on $u(r)$ is sensitive to 
the assumed metal abundance in the model. 
An increase of the $Z/X$ parameter by 10\% has a similar effect on
the sound speed in the outer part of the radiative interior as a 6\% 
increase of age. 

The implication about the age based on $\delta u$ depends also on other
ingredients of the solar model construction such as opacity, nuclear reaction
rates and diffusion coefficients. We will not consider all these effects in 
detail. In Fig. 2 we show few examples of the difference in $u$ between 
models calculated assuming $t=t_{\rm seis}$. Model JCD 
(Christensen -Dalsgaard et al., 1996) is the closest to the 
sun.
The improvement in the opacity data spoils this good agreement. However, as
the comparison with Model 3 shows, the difference in opacity does not 
explain the whole difference between JCD and model 0. We suspect that the 
remaining difference in $u$ may be caused by the difference in the treatment 
of the element settling. The difference between the model 
denoted FR97 (Ciacio et al., 1997) 
and model 0 in the outer
part of the radiative interior is very small. 
A comparison of the plots in  Figs.1 and 2  
shows that the revision the OPAL has 
resulted in changes of $u$ similar to lowering $Z/X$ by 6\%. Thus,
with earlier OPAL opacities we will get solar age lower by 3.6\% 
(0.16 Gy).

In all the cases, values of $\delta u/u$ in the outer part of the 
radiative interior point to 
$t_{\rm seis}>t_{\rm met}$. The difference is model dependent. We will 
quantify it in section 3.1. Finally, let us point out that the result of
inversion shown in Figs 1 and 2 
looks very similar to that of Brun et al. (1998) 
 except for $r<0.1$. The implication concerning the solar age based 
 on $\delta u$ from their inversion would therefore be similar to ours.

\begin{figure}
\epsfxsize=0.99\hsize
\epsffile{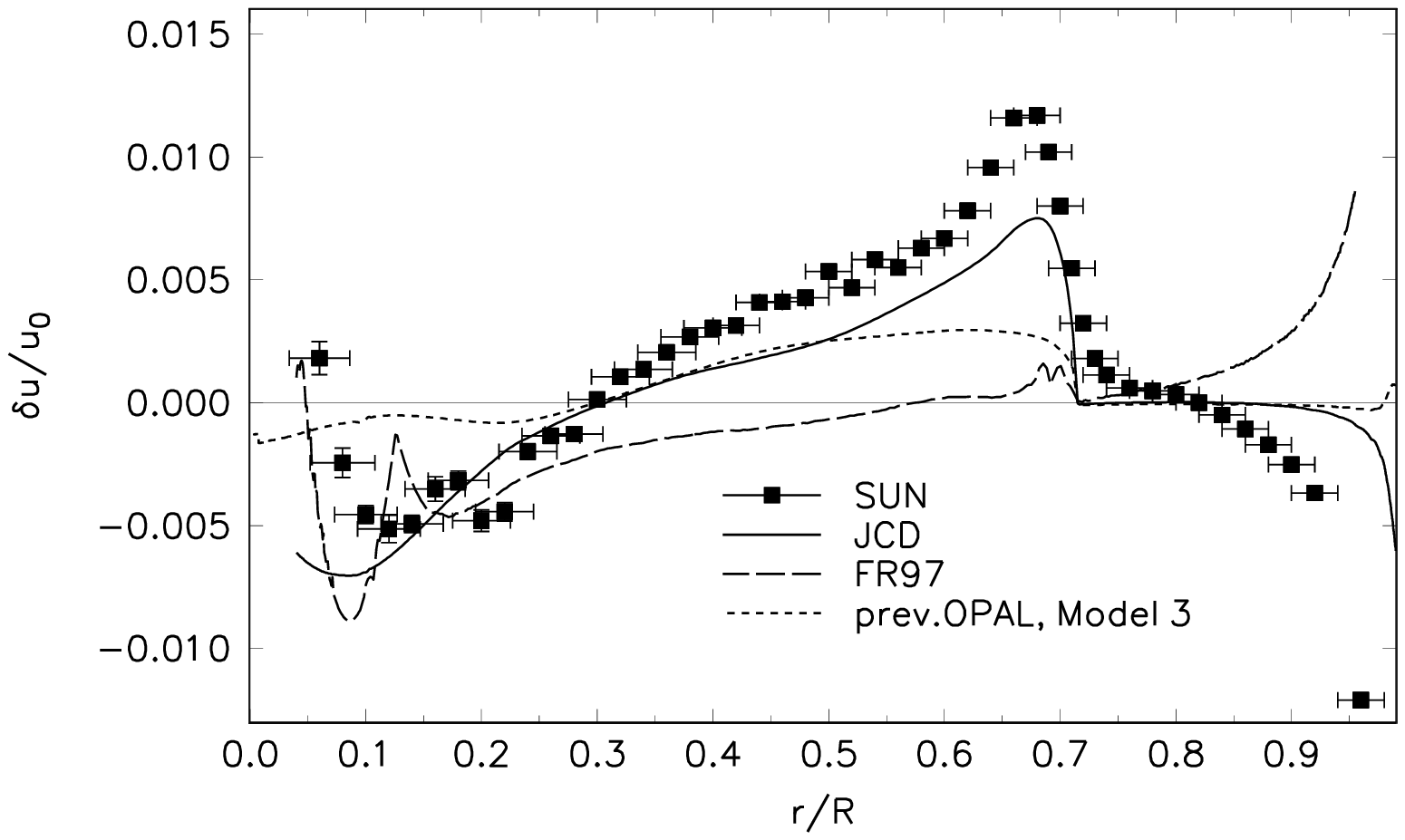}

\noindent Fig.2.  
Relative differences in $u$ between the sun 
and Model 0 determined by means of helioseismic inversion are
compared with the differences 
between different models and Model 0. Model JCD (Christensen-Dalsgaard et al.,
 1995) and Model 3 were calculated with OPAL92 and that denoted FR97 
 with OPAL96 opacities.
\end{figure}

\subsection{Helium abundance}

The value of $Y_{\rm ph}=Y_{\rm seis}$ as determined from the same data 
and with the same reference model is 0.249. It is by 0.006 larger than in 
our standard
model and by 0.010 larger than in model with age 5 Gy. The age inferred from
$Y_{\rm seis}$ would be about 4 Gy. The number is in a reasonable agreement
with  Weiss \& Schlattl (1988). Clearly, there  are conflicting conclusions 
about $t_{\rm seis}$ from
$u(r)$ and $Y_{\rm seis}$. Not surprisingly  Weiss \& Schlattl (1988) 
find rather large minimum values of
$\chi^2$ in their multi-parameter fits.

Adopting higher $Z/X$ values allows to reduce the contradiction. We see in
Table 1 that in model with $Z/X=0.027$, $Y _{\rm ph}$ is close to 
$Y_{\rm seis}$, as well 
as, in Figure 1 we see that $u(r)$ is closer to one inferred by the inversion.
A similar, though smaller, effect is obtained by adopting the previous 
version of the OPAL opacities.  
Still, the most significant difference in $u$ in the outermost part of the
radiative interior cannot be removed by higher $Z/X=0.027$. Modification in 
opacity is an option but it must be quite different than a return to earlier 
version of OPAL. Gough et al. (1996) suggested that the spike of 
$\delta u/u$ at $r\approx 0.68R$ may be a consequence of neglecting a 
macroscopic mixing below the base of convective zone in the standard solar 
models. Models including this effect have been constructed by 
Richard et al. (1996). Such models explain the deficit of Li abundance in the
sun's photosphere and yield better agreement with seismic determination of 
$u$ near the base of convective zone. The effects leads also to an increase of
$Y$ in the envelope.
Macroscopic mixing is a hypothetical effect and its description involves free 
parameters therefore it is not included in the standard models. The effect 
most likely takes place. For present application this means that $Y_{\rm seis}$
and $u$ in the outer part of the envelope is not a safe probe of the 
solar age. In addition, there are difficulties to estimate uncertainties in 
seismic determination of $Y$ following from inadequacies in the  
thermodynamical parameters.

\subsection{Estimates of $t_{\rm seis}$ based on selected values of $u$ and
$Y _{\rm ph}$}

For the sake of illustration of the 
discrepancies we will give estimates of $t_{\rm seis}$ based on 
different observables. 
Unlike Weiss \& Shlattl (1998), we will not try to fit
simultaneously more than one parameter because our aim is only to quantify the
problems with the assessment of the solar age with the method reviewed in this
section. Furthermore, the meaning of the formal $\chi^2$-minimalization 
procedure is problematical in present case, as in fact Weiss \& Shlattl 
(1998) emphasized.

In Table 2 we provide a list of the selected observables, $Q$ with errors, 
with its estimated $1\sigma$ uncertainty $\Delta Q/Q$, and the 
quantity
\begin{equation}
\alpha _{Q}= {d\ln Q\over d\ln(t/t_{\rm met})0}  \quad ,  
\end{equation}

\noindent which measures sensitivity of each observable to the solar age.
The values of $\tilde u =uR/GM$ and $Y _{\rm ph}$ are from the inversion 
described in subsection 3.1. The estimates of uncertainties, $\Delta Q/Q$, 
are from Degl'Innocenti et al. (1997).

\begin{table}

\caption{Selected seismic observables and their 
$1\sigma$ uncertainties , $\Delta Q/Q$.} 

\begin{tabular}{cccc}
\hline
  $Q$      & $\alpha_Q$  &  $Q_\odot$ & $\Delta Q/Q$ \\
\hline
\~u$(0.3)$&  0.03& 0.4782  &  $\pm$0.1\%\\
\~u$(0.4)$&  0.05& 0.3618  &  $\pm$0.1\% \\
\~u$(0.5)$&  0.07& 0.2820  &  $\pm$0.12\% \\
\~u$(0.6)$&  0.09& 0.2218  &  $\pm$0.14\% \\
\~u$(0.65)$& 0.08& 0.1952  &  $\pm$0.14\% \\
$Y_{ph}$&  -0.20 & 0.249   & $\pm1.4$\% \\
\hline
\end{tabular}
\label{Tab2}
\end{table}

In Table 2 we list the values of the selected observables 
calculated in  
the three standard solar models. 
\begin{table}
\caption{ Values of $\tilde u$ and $Y_{\rm ph}$}
\begin{tabular}{lcccc}
\hline
$Q_i$ &JCD & model 0 & FR97 \\
\hline
\~u$(0.3)$& 0.4781  & 0.4781 & 0.4772   \\
\~u$(0.4)$& 0.3612  & 0.3607 & 0.3603   \\
\~u$(0.5)$& 0.2812  & 0.2805 & 0.2803   \\
\~u$(0.6)$& 0.2214  & 0.2203 & 0.2204   \\
\~u$(0.65)$& 0.1945  & 0.1932 & 0.1932   \\
$Y_{ph}$  & 0.245   & 0.243  & 0.238    \\
\hline
\end{tabular}
\end{table}

\begin{table}
\caption{Helioseismic estimate of solar age (Gy), as inferred from 
 the differences $Q-Q_i$, calculated for different SSMs.}
\begin{tabular}{lcccc}
\hline
$Q_i$ &JCD & model 0 & FR97 \\
\hline
\~u$(0.3)$& $4.60 \pm 0.15$  & $4.60 \pm 0.15$ &  $4.90 \pm 0.14 $ \\
\~u$(0.4)$& $4.72 \pm 0.09$  & $4.86 \pm 0.10$ &  $4.96 \pm 0.10 $  \\
\~u$(0.5)$& $4.76 \pm 0.08$  & $4.93 \pm 0.08$ &  $4.98 \pm 0.08 $  \\
\~u$(0.6)$& $4.66 \pm 0.07$  & $4.93 \pm 0.08$ &  $4.90 \pm 0.08 $  \\
\~u$(0.65)$& $4.78 \pm 0.08$  & $5.20 \pm 0.09$ & $ 5.20 \pm 0.09 $ \\
$Y_{ph}$   & $4.21 \pm 0.29$  & $4.04 \pm 0.28$ & $ 3.64 \pm 0.25 $ \\
\hline
\end{tabular}
\end{table}

In Table 4 we provide the values of $t$ inferred from the differences 
between the sun and the  models by using the various observables $Q$. 
The numbers mostly quantify only the effects discussed earlier in this section.

\section {Direct and almost direct use of measured frequencies}

It is unfortunate that the parameters of seismic models which exhibit greatest
sensitivity to solar age are, for various reasons, unreliable. The sound 
speed in the inner core cannot be precisely measured because the 
inversion is not 
accurate enough. Other parameters are formally very accurate but 
we cannot trust model predictions. Since the nature of 
the uncertainties is so diversified, we are reluctant to quote any 
quantity as a best value of $t_{\rm seis}$ and its errors.    

Choosing, instead, a direct use of frequency differences we face a  
another problem. The formal approach to determination of $t_{\rm seis}$ is
minimization of
\begin{equation} 
\chi^2={1\over J}\sum_{j=1}^J
\left({\nu_\odot-\nu_{\rm model}(t)\over\sigma}\right)_j^2,
\end{equation} 
where in the sum includes all $J=1945$ p-modes in the set, and 
$\sigma$ are measurements error. The problem is revealed in
Fig.2  where we may see that $\chi^2$ 
depends only very weakly on the age. There is a minimum near 5.2 Gy, but it 
is
very shallow and  does not allow a trustworthy estimate of $t_{\rm seis}$. 

\begin{figure}
\epsfxsize=0.9\hsize
\epsffile{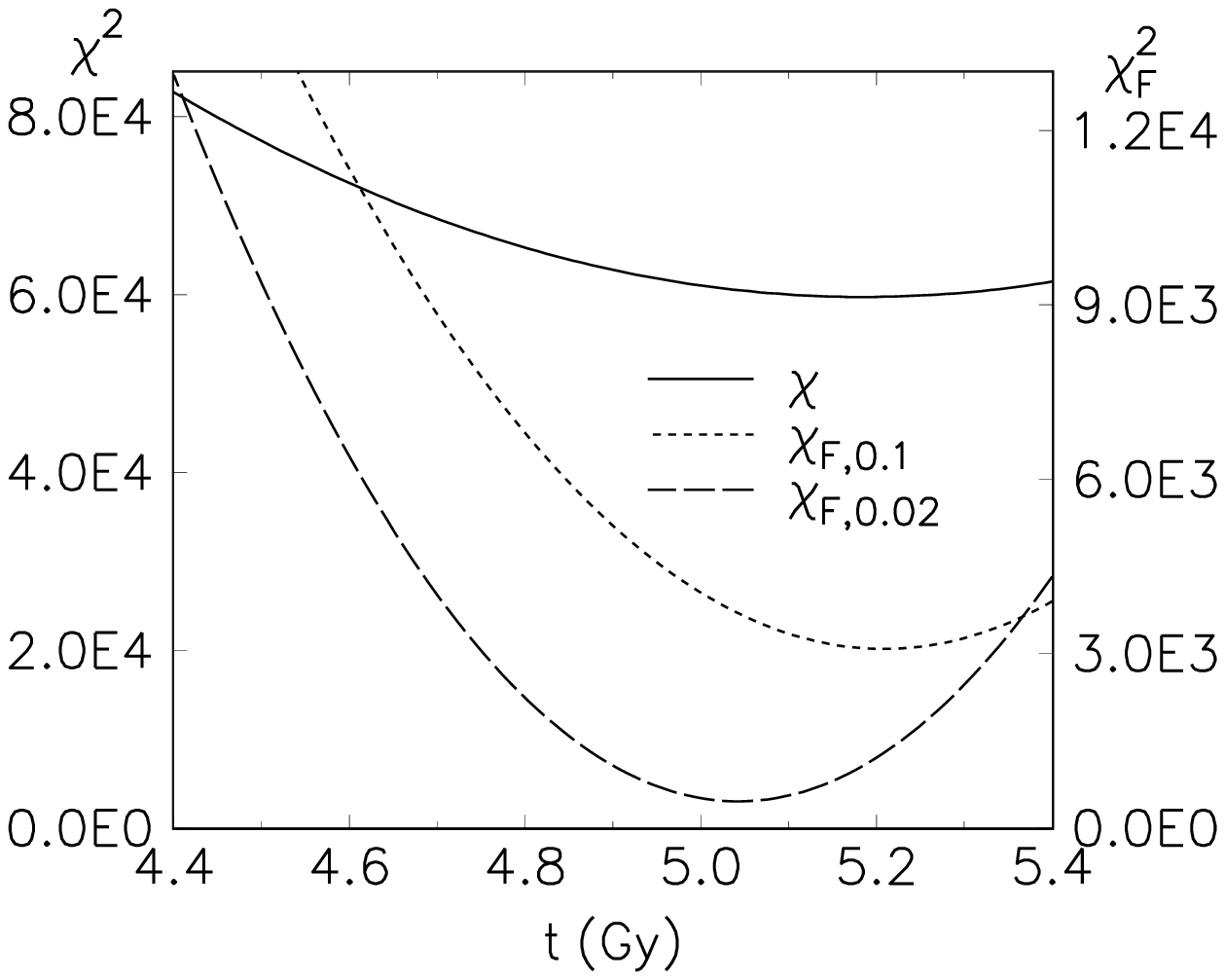}

\noindent Fig.3.
Determination of the solar age by fitting p-mode frequencies. 
Values of $\chi^2$ (left vertical axis, solid line) are calculated with
Eq. 2. Values of $\chi_{F,s}^2$ for $s=0.1$ and $s=0.02$
(right vertical axis, dashed lines) are calculated also with Eq. 2, but
with $\nu_\odot$ replaced $\nu_F$ (see Eq. 3). The choice $s=0.1$ implies 
use of all p-mode frequencies and elimination of the near surface differences
between the sun and the model. With $s=0.02$ we use only modes with the lower
turning point above $0.8R$ and we additionally eliminate effects of 
inadequacies in the treatment of convection. 

\end{figure}

This problem 
is a consequence of the fact that the main part of the frequency differences 
between 
the sun and the model has nothing to do with the differences in the internal
structure but rather it is caused by inadequacies in the treatment of 
oscillations
in the outer layers, where neglect of nonadiabatic effects and dynamical 
effects
of convection is not justified. These inadequacies are significant in 
outermost layers above $r=0.99R$ i.e. above the lower turning point of 
all the p-modes in the set. The lower turning point is determined by 
the parameter $(\ell+0.5)/\nu$. Its maximum value for modes in our 
set is 0.1 and corresponds to the turning point $r_t=0.99R$.
Sufficiently far above the turning point the 
relevant eigenfunctions, except for normalization, are $\ell$-independent.
Therefore, we
may expect that the part of the frequency difference due to the effects in
in the layers above $r=0.99R$ scales as $F(\nu)/I_j$, where $I_j$ is the mode
inertia calculated upon assuming the same normalization of the eigenfunctions 
in the photosphere. 

In order to eliminate these near-surface contamination, we fitted 
$F(\nu)$ in a 
polynomial form to the frequency differences 
$\nu_{\odot,j}-\nu_{{\rm model},j}$ and  
considered only the residual part of the differences
\begin{equation}
\nu_{F,j}= \nu_{\odot,j}- {F(\nu_{\odot,j})\over I_j}.
\end{equation}
The quantity $\nu_{F,j}-\nu_{{\rm model},j}$ is the part of the frequency
difference that may be attributed only to the difference in the internal 
structure. 
In Figure 2 we plot two $\chi_{F,s}^2(t)$ functions, which is a 
modified $\chi^2$ with
$\nu_\odot$ replaced $\nu_F$. The parameter $s$ is the maximum  
value of the quantity $(\ell+0.5)/\nu$ ($\nu$ in $\mu$Hz), which determines
the lower turning point, 
allowed in the set of modes. The case $s=0.1$ corresponds to including all 
1945 p-modes. The case $s=0.02$ corresponds to a truncated set which includes 
only 956 modes with $r_t<0.8R$ 
In the latter case, we additionally remove effects 
of inadequate treatment of convection which are responsible for large values 
of $\delta u/u$ above 0.9. The minima of the modified $\chi$ are pronounced 
and therefore we may, at least formally, determine solar age and its 
uncertainty.
Not surprisingly, the minimum is 
deeper for $s=0.02$.  Still, the minimum value is $\gg 1$. One may see
in Fig. 1 that  $\delta u/u$ in the radiative interior
cannot be compensated by an adjustment of the age. 

In Table 5, we list the values of $t_{\rm seis}$ determined as minima of 
$\chi_{F,0.1}$ and $\chi_{F,0.02}$. The errors are determined as the distances
from $t_{\rm seis}$, where $\chi^2=2\chi^2(t_{\rm seis})$.

\begin{table}
\caption{Seismic age from p-mode frequencies}
\begin{flushleft}
\begin{tabular}{ccccc}
\hline
 & $s=0.1$ & & $s=0.02$ & \\
 $Z/X$  & $t_{\rm seis}$ & $\chi^2$ & $t_{\rm seis}$ &  $\chi^2$\\
 \hline
0.0245 & $5.22\pm0.40$ & $3.16\times10^3$ & $5.04\pm0.13$ & $5.2\times10^2$\\
0.0270 & $4.91\pm0.34$ & $3.45\times10^3$ & $4.77\pm0.13$ & $5.3\times10^2$\\
\hline
\hline
\end{tabular}
\end{flushleft}
\end{table}

The results shown in Table 5 are consistent with implications from 
$\delta u(r)$ discussed in the previous section. There are only few modes 
sensitive to $u$ in the inner core, where $\delta u$ is not consistent 
with high $t_{\rm seis}$. Also, even with $s=0.1$ there are not many 
modes sensitive to $Y _{\rm ph}$. The results agree with those of  
Weiss \& Schlattl (1988). All this does 
not mean that we should treat $t_{\rm seis}$ given in Table 5 as realistic 
estimates of solar age. Rather, we think, the high values obtained for models  
with the standard metal abundance reflect an attempt to compensate such 
deficiencies of the model as too low opacity and/or neglect of macroscopic 
mixing beneath the base of convective envelope.
With $Z/X=0.027$ we obtained $t_{\rm seis}$ which still higher but, 
within the errors, consistent with $t_{\rm met}$.

\section{Solar age from small separations}

\begin{figure}
\epsfxsize=0.9\hsize
\epsffile{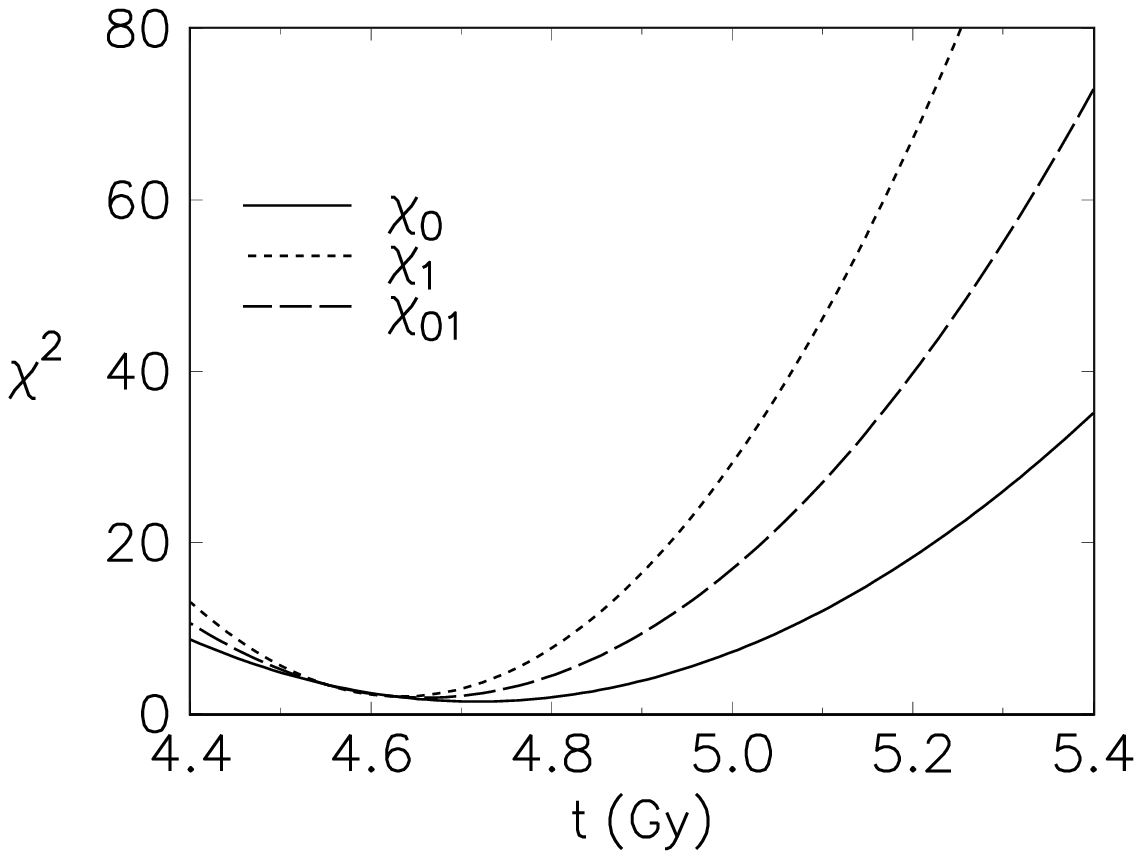}

\noindent Fig. 4.
Determination of the solar age by fitting small frequency separations 
(see Eq. 3). The quantity $\chi_0$ is defined in Eq. 4,  
$\chi_1$ and $\chi_{0,1}$ are defined immediately after. 

\end{figure}

\begin{table*}
\caption{Seismic age from small separations}
\begin{flushleft}
\begin{tabular}{ccccccc}
\hline
  & $\ell=0$ & & $\ell=1$ & & $\ell=0\& 1$ & \\
  $Z/X$ & $t_{\rm seis}$ & $\chi_0^2$ & $t_{\rm seis}$ & $\chi_1^2$& 
  $t_{\rm seis}$ & $\chi_{01}^2$\\
 \hline
0.0245 & $4.71\pm0.14$ & 1.40 & $4.64\pm0.08$ & 1.34& $4.66\pm0.11$ & 1.52\\
0.0270 & $4.63\pm0.14$ & 1.48 & $4.54\pm0.08$ & 1.44& $4.57\pm0.11$ & 1.68\\
\hline
\hline
\end{tabular}
\end{flushleft}
\end{table*}

The inner core is the region where the sound speed is most sensitive 
to the age.
Inversion for $u$ in this region is unreliable but this does not mean that
oscillation frequencies are not affected by the sound speed modifications 
near the center.
The quantities which are most sensitive to changes in the inner core are
small separations 
\begin{equation}
D_{\ell,n}=\nu_{\ell,n}-\nu_{\ell+2,n-1}
\end{equation}

\noindent for $\ell=0$ and 1.
In fact it has been recognized long time ago that data on 
$D_{\ell,n}$ may be used for measuring stellar ages (Ulrich, 1986; 
Christensen-Dalsgaard, 1988; Gough \& Novotny, 1990). 
  
In our set we have data on $D_{0,n}$ for $n$ from 10 to 32 
and on $D_{1,n}$ for $n$ from 10 to 27. We now form three 
age indicators, 
\begin{equation}
\chi_0^2={1\over23}\sum_{n=10}^{32}{(D_{0,n,\odot}-D_{0,n,{\rm model}}(t))^2
\over\sigma_{0,n}^2+\sigma_{2,n-1}^2},
\end{equation}
$\chi_1^2$, which is defined is the same way as $\chi_0^2$
but for $\ell=1$, and $\chi_{01}^2$, which includes small separations 
both for $\ell=0$ and  $\ell=1$. 

The behavior of the three indicators 
is shown in Fig.4. The $\chi^2$ minima occur now at the ages which 
are only somewhat larger than $t_{\rm met}$ and have values only somewhat 
higher than 1. 
Table 6 summarizes information about the minima for models with the standard 
and the enhanced value of $Z/X$. In the latter case the minima occur still 
closer to $t_{\rm met}$, but the difference is small and  cannot be regarded as
significant. The ages $t>5$ Gy are clearly disfavored.
There is a rough agreement of our result with that of 
Guenther \& Demarque (1997),
who relied on comparison of frequencies for $\ell$ up to 100 and small 
separations for $\ell$ up to 10. Also in their comparisons the strong 
case for $t_{\rm seis}\approx t_{\rm met}$ comes from small separations
at $\ell=0$ and 1.
 
We believe that only in the case of inference based on the small 
separations it is justified to speak about ``age determination" because only 
with these observables we attain $\chi^2 \sim 1$. Furthermore, only
in this case the inference is truly robust to other uncertainties still 
present in the standard model construction.
The over-all uncertainty of the seismic 
measurement of the solar age with the data on small separations is 
not significantly larger than the formal errors quoted in Table 6.
The effect of the $Z/X$ uncertainty, as we may see in this table, is 
$\le0.1$ Gy. Now we will review other uncertainties that may affect 
small separations.

Effect of  uncertainties on the age indicators $\chi_0$,$\chi_1$,
and $\chi_{01}$ are may be asses from data in Table 7. 
The effect of the opacity is revealed by comparison of models 0 with 
models 3 and JCD and we may see that it is small. As we discussed in Section 3,
the difference in opacity does not explain the whole difference 
in the sound speed between 
the models 0 and JCD. We alluded that the treatment of the element settling 
may contribute. In any case the implication for $t_{\rm seis}$ are 
certainly within the uncertainties quoted in Table 6.
We should note that JCD model which is characterized by the lowest value of
$\chi_{0,1}^2$ yields also the values of $t_{\rm seis}$ which are the 
closest to $t_{\rm met}$ on the basis of the seismic observables
listed in Table 4.

Ignoring gravitational settling altogether (see Model 5 in Table 7) has 
a significant effect on small separations. However, the effect is now part of 
the physics included in the standard modeling of the sun. 

Calculated values of the small separations are affected by the 
nuclear reactions cross-sections. The most important effect is expected
from changes in the branching ratio of the $^3$He+$^4$He 
to the $^3$He+$^3$He reaction. Its increase implies more neutrino energy
losses, less economic hydrogen burning, and consequently less Hydrogen in
the center of the sun. Such models mimic ones with $t>t_{\rm seis}$. 
However with currently adopted uncertainties in the cross section 
(see Model 4 in Table 7) the consequences for the age indicators are 
not significant.
 
Mixing of Hydrogen and Helium reduces the $\mu$-gradient in the 
core and thus has a similar effect as a lower age. This is not a standard
effect and we feel that there is not enough justification to 
consider it as a source of uncertainty. Certainly macroscopic mixing 
at the base of the convective zone is of more concern because we have some
evidence for it. The mixing affects gravitational settling and therefore may
have an appreciable effect on small separations.

\begin{table}
\caption{Seismic age indicators from small separations 
in various models at $t=4.57$ Gy and 
 $Z/X$=0.0245}
\begin{flushleft}
\begin{tabular}{cccc}
\hline
 MODEL & $\chi_0^2$ & $\chi_1^2$  & $\chi_{0,1}^2$ \\
 \hline
0 &  2.89 & 2.32  & 2.64 \\
3 &  2.31 & 1.69 & 2.04 \\
JCD & 1.82 & 0.99  & 1.46 \\
4 &  2.75 & 2.01  & 2.43 \\
5 &  20.06 & 44.84 & 30.94 \\
\hline
\hline
\end{tabular}
\end{flushleft}
Model 4 is the same as model 0, but with a 3.2\% increase of the $^3$He+$^4$He 
reaction cross-section and a 6\% decrease of the $^3$He+$^3$He 
reaction cross-section. Model 5 is the same as Model 0 but ignoring the
effect of gravitational settling.  
\end{table}

Small separations are influenced by the centrifugal and magnetic distortion
(Dziembowski \& Goupil, 1998). The effect of centrifugal distortion in the 
sun is small because it is a very slowly rotating star. However, at a 
rotation rate five time higher the values of $D_{0,n}$ for $n\approx20$ are 
reduced by $\approx0.5$ $\mu$Hz, which corresponds to about 0.5 Gy. Thus
the effect has to be kept in mind when we will have small separations data 
for other stars. For the sun, the magnetic effect in high activity years are
significant but they may easily be purged (Dziembowski \& Goode, 1997).
The problem does not concern the frequencies used in present paper because we 
used data from 1996/97 season when solar activity was at its minimum.

\section{Conclusions}

There is no evidence from helioseismology that the solar age is different
from $t_{\rm met}$. While it is true that with models adopting higher values 
one may achieve a better agreement for most of seismic data this cannot be
regarded as an argument that the sun is older than the meteoritic data 
indicate. What seismic testing of current standard solar models reveals is
the need for increase of the sound speed in the outer part of the radiative
zone by about half percent ($\sim 0.01$ in $u$). The required change 
may be partially achieved by an age increase above 5 Gy but it also may be
caused by an increase of opacity in the relevant region.
The required opacity increase may result from some still ignored effects
in the OPAL calculations but also may indicate that the metal content in the
outer part of the radiative zone is higher. We showed that adopting 
$Z/X=0.027$, which is by 10\% higher than the standard one but still  
within the error bars of determination, we derive $t_{\rm seis}$ below 5 Gy 
and only marginally inconsistent with $t_{\rm met}$.

In fact, helioseismology provides a strong support for the assumption
$t=t_{\rm met}$.
We showed that small frequency separations $\nu_{0,n}-\nu_{2,n-1}$ 
and $\nu_{1,n}-\nu_{3,n-1}$ determined from the data are in a drastic 
disagreement with the models older than 4.9 Gy and 
they are in a good 
agreement with the models calculated assuming $t=t_{\rm met}$.
The age of the sun  determined from the best seismic data and with use 
of our standard models, which were calculated with the 
latest OPAL opacity data and the standard metallicity 
parameter $Z/X=0.0245$, is
$$t_{\rm seis}=4.66\pm0.11 \, {\rm Gy}$$
Outside the error range $\chi^2>2\chi_{\rm min} ^2$.

The small separations are only weakly affected by uncertainties 
in the opacity. Still, models with enhanced opacity in the outer
part of the radiative zone yield values of $t_{\rm seis}$ even closer 
to $t_{\rm met}$. We, thus, conclude that the inadequacies of the 
current solar models cannot be reconciled by departing from 
standard assumption about solar age but the resolution must be searched in
opacity enhancement.

Our answer to the question  how 
accurately we can determine age of the sun using  
stellar evolution theory and helioseismic data, posed by Paczy\'nski (1997),
is more optimistic than the answer of  Weiss \& Schlattl (1988). 

The error bars given above may be somewhat underestimated. 
Taking into account the
uncertainties beyond those included in the formal errors, the accuracy of 
the astrophysical estimate the solar age, is in our opinion, 
$\sim 0.2$ Gy or  4\%, which is significantly 
better than 0.5 Gy, as suggested by  Weiss \& Schlattl (1988). 

The cause for the discrepant 
estimates is in the use of different observables. We believe that only  
the small separations are good probes of the solar age based on p-mode
frequency data. 
Others, like frequencies themselves, seismically inferred sound speed, 
and photospheric Helium abundance  are too sensitive 
to the opacity to be regarded as a reliable tools for measuring solar and
stellar ages. 

We examined various effects that may contribute to the uncertainty of 
the age determination from the small separations. None of the 
uncertainties in the physics included in modern standard solar 
models was found very significant. However, we identified few effects beyond 
standard model that may have large effect on the small separations. 
Perhaps most 
important is a macroscopic mixing in the outermost part of the radiative 
interior. We considered also the effects of the centrifugal and magnetic 
forces and we pointed out that while they are not important for
our seismic estimate of the solar age they must be kept in mind when 
in interpretation of data on the small separations from years of high 
magnetic activity as well as the data for stars rotating more rapidly 
than the sun. 

All the seismic observables we discussed here are still available
only for the sun. The observables that we are likely to have in not
too distant future for other stars are the small separations. Measuring these 
parameters is one of the main goals of the three currently prepared or 
planned space 
asteroseismic missions. It is very fortunate that, as we have shown, the 
small separations are the best seismic age indicators derived from p-mode 
frequencies. There is a potential for measuring stellar ages based on g-modes,
which are excited in a number of stars. However, also in this case it is 
essential to check robustness of the seismic dating. 

\begin{acknowledgements}

B.R. and W.A.D thank   
Copernicus Astronomical  Center in Warsaw
and Istituto Nazionale di Fisica Nucleare, respectively, for 
the hospitality. 
This research was supported by Ministero
dell'Universit\`a e Ricerca Scientifica (MURST)
through grant 40\% 1998 and by the Polish grant KBN-2P03D-014.
\end{acknowledgements}
%

\end{document}